\documentclass[aps,prd,preprint,tightenlines,groupedaddress,nofootinbib,showpacs,byrevtex]{revtex4}
\usepackage{amssymb,latexsym}
\usepackage{amsmath,amsbsy}
\usepackage{epsfig,bm}
\usepackage{graphicx,comment}
\DeclareMathOperator{\tr}{tr}

\begin{document}
\def\a{{\alpha}}
\def\b{{\beta}}
\def\d{{\delta}}
\def\D{{\Delta}}
\def\e{{\varepsilon}}
\def\g{{\gamma}}
\def\G{{\Gamma}}
\def\k{{\kappa}}
\def\l{{\lambda}}
\def\L{{\Lambda}}
\def\m{{\mu}}
\def\n{{\nu}}
\def\o{{\omega}}
\def\O{{\Omega}}
\def\S{{\Sigma}}
\def\s{{\sigma}}
\def\th{{\theta}}

\def\ol#1{{\overline{#1}}}

\def\Dslash{D\hskip-0.65em /}

\def\CPT{{$\chi$PT}}
\def\QCPT{{Q$\chi$PT}}
\def\PQCPT{{PQ$\chi$PT}}
\def\tr{\text{tr}}
\def\str{\text{str}}
\def\diag{\text{diag}}
\def\order{{\mathcal O}}

\def\cC{{\mathcal C}}
\def\cB{{\mathcal B}}
\def\cT{{\mathcal T}}
\def\cQ{{\mathcal Q}}
\def\cL{{\mathcal L}}
\def\cO{{\mathcal O}}
\def\cA{{\mathcal A}}
\def\cQ{{\mathcal Q}}
\def\cR{{\mathcal R}}
\def\cH{{\mathcal H}}
\def\cW{{\mathcal W}}
\def\cM{{\mathcal M}}

\def\eqref#1{{(\ref{#1})}}

 
\title{Extrapolations of Lattice Meson Form Factors}
\author{T.~B.~Bunton}
\email[]{bunton@phy.duke.edu}
\author{F.-J.~Jiang}
\email[]{fjjiang@phy.duke.edu}
\author{B.~C.~Tiburzi}
\email[]{bctiburz@phy.duke.edu}
\affiliation{Department of Physics\\
Duke University\\
P.O.~Box 90305\\
Durham, NC 27708-0305}

\date{\today}


\begin{abstract}
We use chiral perturbation theory to 
study the extrapolations necessary to make physical predictions 
from lattice QCD data for the electromagnetic form factors of pseudoscalar mesons. 
We focus on the quark mass, momentum, lattice spacing, and volume dependence
and apply our results to 
simulations employing mixed actions of Ginsparg-Wilson
valence quarks and staggered sea quarks.
To determine charge radii at quark masses on the lattices 
currently used, we find that all extrapolations 
except the one to infinite volume 
make significant contributions to the systematic error.
\end{abstract}

\maketitle

\section{Introduction}

As computing resources and numerical algorithms improve, first principles determination of 
hadronic observables will be possible with lattice QCD. In the foreseeable future, these numerical 
determinations will rely on effective field theories to address systematic errors in lattice 
data.
There has been considerable effort to understand and compute effects from the finite
volume of the lattice, the discretization chosen for fermions, and most notably the quark mass dependence 
of observables. 
To address systematic error related to the treatment of the fermionic determinant, quenched
chiral perturbation theory~%
\cite{Morel:1987xk,Sharpe:1992ft,Bernard:1992mk}, 
and partially quenched chiral perturbation theory%
~\cite{Bernard:1994sv,Sharpe:1997by,Golterman:1998st,Sharpe:2000bc,Sharpe:2001fh}
have been developed. In the respective approximation made in the former case, 
the determinant is replaced by a constant leading to uncontrolled systematic error, while in the 
latter the determinant is computed but with larger quark masses used than in the propagators 
connected to external legs.  Only through this latter  
approximation can connection to real QCD observables be made; moreover, effective 
field theory is required to make this connection systematically.

In this work we study the extraction of the 
electromagnetic charge radii of 
pseudoscalar mesons from  lattice QCD 
within the framework of partially quenched chiral perturbation theory.
We consider the various extrapolations in quark mass, momentum, volume
and lattice spacing needed to extract the radii.  
Experimentally, the pion charge radius is rather well determined 
from pion scattering off atomic electrons~\cite{Amendolia:1986wj,Eidelman:2004wy}.
As such it can be used as a crucial test of lattice and effective field theory methods.
Away from small momentum transfer, the pion form factor
has been experimentally probed from the virtual pion cloud of the nucleon~\cite{Volmer:2000ek}; 
however, extraction is limited by model dependence assumed in extrapolating the experimental data to zero virtuality. 
Ultimately lattice methods will enable first principles QCD calculation of  
meson form factors over a wide range of momentum transfer.

The original lattice QCD calculations of the pion form factor were pursued by two groups%
~\cite{Martinelli:1988bh,Draper:1989bp}. Since these pioneering 
calculations
there have been various further computations and refinements using improved actions, larger volumes, 
and different lattice fermions; for recent investigations see, e.g.,%
~\cite{Nemoto:2003ng,vanderHeide:2003kh,Abdel-Rehim:2004gx,Capitani:2005ce}. 
These calculations are all limited by the quenched approximation. 
There have been, however, recent lattice calculations that include dynamical quarks%
~\cite{Bonnet:2004fr,Hashimoto:2005am}. 
The results with dynamical quarks in~\cite{Bonnet:2004fr} use a  mixed lattice action of 
domain wall valence quarks on staggered sea quarks, and the lightest pion mass $\sim 300 \, \texttt{MeV}$
is arguably within the chiral regime. Indeed such mixed action 
simulations are currently popular due to both the publicly available MILC configurations~\cite{Bernard:2001av},
and the desirable chiral symmetry properties of domain wall fermions~\cite{Kaplan:1992bt} (or more generally, of Ginsparg-Wilson 
fermions~\cite{Luscher:1998pq}). 
For other observables calculated with such mixed actions, see%
~\cite{Renner:2004ck,Bowler:2004hs,Beane:2005rj,Edwards:2005kw,Alexandrou:2005em,%
Edwards:2005ym,Beane:2006mx,Beane:2006pt,Beane:2006fk,Beane:2006kx}.
The low-energy effective theory for mixed lattice actions is a partially quenched chiral perturbation theory
even when the valence and sea quark masses are degenerate. An additional
reason for us to study the pion form factor is to learn about 
the low-energy constants of mixed action chiral perturbation theory~\cite{Bar:2005tu}. 
At the order we work, only one new parameter $C_\text{mix}$ [see Eq.~\eqref{eq:mixed}] 
enters in the continuum extrapolation of the pion charge radius. 
We find, however, that the data in~\cite{Bonnet:2004fr} allow for only a rough estimate
of the parameter $C_\text{mix}$, a problem that can be remedied with more lattice 
data for differing valence quark masses, or at different lattice spacings.

The organization of the paper is as follows. 
First in Sec.~\ref{s:pqqcd}, we review the basics
or partially quenched chiral perturbation theory. 
In Sec.~\ref{s:infinite}, we derive the meson 
form factors at one-loop order in partially quenched 
chiral perturbation theory. Our result improves upon
an earlier calculation~\cite{Arndt:2003ww} by using a computationally 
judicious choice for the quark charges. 
Next in Sec.~\ref{s:a}, we include the effects 
from the lattice discretization for the case
of a mixed action of Ginsparg-Wilson valence quarks
and staggered sea quarks. The finite volume corrections
are presented in Sec.~\ref{s:finite}. Extrapolations 
are considered in Sec.~\ref{s:extrap}, where 
we use our results to investigate the chiral, momentum, volume, and continuum 
extrapolations of meson form factors. 
Lastly we conclude with a brief summary of our work, Sec.~\ref{s:summy}.

\section{Partially Quenched Chiral Lagrangian} \label{s:pqqcd}

In partially quenched QCD, the quark part of the continuum Lagrangian is written as%
\begin{eqnarray}\label{eqn:LPQQCD}
  {\cal L}
  &=&
  \sum_{j=1}^{9}
  \bar{Q}_j(\Dslash + m_Q) Q_j \,
.\end{eqnarray}
The nine quarks appear in the vector 
\begin{equation}
  Q=(u,d,s,j,l,r,\tilde{u},\tilde{d},\tilde{s})^{\text{T}}
,\end{equation}
that transforms in the fundamental representation of
the graded group $SU(6|3)$%
~\cite{
BahaBalantekin:1981qy,BahaBalantekin:1982bk}.
The quark mass matrix is given by 
\begin{equation}
  m_Q=\text{diag}(m_u,m_d,m_s,m_j,m_l,m_r,m_u,m_d,m_s),
\end{equation}
to maintain the cancellation of path integral determinants from 
the valence and ghost sectors.
Effects of dynamical quarks are
present due to the contribution of the finite-mass 
sea quarks. 
Additionally we choose to work in the isospin limit
in the valence and sea sectors: $m_d = m_u$ and $m_l = m_j$. 

The light quark electric charge matrix $\cQ$ is not uniquely
defined in partially quenched QCD~\cite{Golterman:2001qj}.  
The only constraint one must impose is that
the charge matrix $\cQ$ has vanishing
supertrace.
Following~\cite{Tiburzi:2004mv,Detmold:2005pt}, we use
\begin{equation}
  \cQ
  =
  \diag
  \left(
    q_u,q_d,q_s,q_j,q_l,q_r,q_u,q_d,q_s
  \right)
,\end{equation}
along with the condition $q_j + q_l + q_r = 0$.
QCD is recovered in the limit of degenerate valence and sea 
quarks only for the particular choice: $q_u = q_j = \frac{2}{3}$, 
and $q_d = q_s = q_l = q_r = - \frac{1}{3}$. Letting the charges 
be arbitrary, however, enables us to track the flow of charge
in loop diagrams.

For massless quarks,
the Lagrangian in Eq.~(\ref{eqn:LPQQCD}) exhibits the graded symmetry
$SU(6|3)_L \otimes SU(6|3)_R \otimes U(1)_V$ that we assume 
is spontaneously broken to $SU(6|3)_V \otimes U(1)_V$. 
The low-energy effective theory of partially quenched QCD is written in terms 
of the pseudo-Goldstone mesons emerging from spontaneous chiral symmetry breaking. 
At lowest order in the chiral expansion, the dynamics of these 
mesons can be described by the $\order(p^2)$ Lagrangian%
\footnote{
Here $p  \sim m_\pi$ where $p$ is an external momentum.
}
\begin{equation}\label{eqn:Lchi}
  {\cal L} =
  \frac{f^2}{8}
    \str\left(D_\mu\Sigma^\dagger D_\mu\Sigma\right)
    - \frac{\l}{4}\,\str\left(m_Q^\dagger\Sigma+m_Q\Sigma^\dagger\right)
    + \a\partial_\mu\Phi_0\partial_\mu\Phi_0
    + \mu_0^2\Phi_0^2
,\end{equation}
where
\begin{equation} \label{eqn:Sigma}
  \Sigma=\exp\left(\frac{2i\Phi}{f}\right)
,\end{equation}
\begin{equation}
  \Phi=
    \left(
      \begin{array}{cc}
        M & \chi^{\dagger} \\ 
        \chi & \tilde{M}
      \end{array}
    \right)
,\end{equation}
$f=132 \, \texttt{MeV}$,
and we have defined the electromagnetic gauge-covariant derivative
$D_\mu\S=\partial_\mu\S+ie\cA_\mu[\cQ,\S]$.
The str() denotes a supertrace over flavor indices.
The $M$, $\tilde{M}$, and $\chi$ are matrices
of pseudo-Goldstone bosons with quantum numbers of $q\ol{q}$ pairs,
pseudo-Goldstone bosons with quantum numbers of 
$\tilde{q}\ol{\tilde{q}}$ pairs, 
and pseudo-Goldstone fermions with quantum numbers of $\tilde{q}\ol{q}$ pairs,
respectively.
Upon expanding the Lagrangian in \eqref{eqn:Lchi} one finds that
quark basis 
mesons with quark content $Q\bar{Q'}$
have masses
\begin{equation}\label{eqn:mqq}
  m_{QQ'}^2=\frac{\lambda}{f^2}(m_Q+m_{Q'})
.\end{equation}

The flavor singlet field appearing above is given by $\Phi_0=\str(\Phi)/\sqrt{6}$.
Just as in chiral perturbation theory, but in contrast to the quenched case, 
the singlet field is rendered heavy by the strong 
$U(1)_A$ anomaly and is integrated out of the theory.
The resulting flavor neutral propagators, however, deviate from 
simple pole forms~\cite{Sharpe:2001fh}. We do not display these
propagators here as they are not explicitly needed in our final results
for meson form factors.

Additionally there are three terms in the $\order(p^4)$ Lagrangian
\begin{eqnarray} \label{eqn:L4PQQCD}
  {\cal L}
  &=&
  \a_4 \,
    \str(D_\mu\Sigma D_\mu\Sigma^\dagger) \,
    \str(m_Q^\dagger\Sigma+m_Q \Sigma^\dagger)
  +
  \a_5 \,
    \str(D_\mu\Sigma D_\mu\Sigma^\dagger
        (m_Q^\dagger\Sigma+m_Q \Sigma^\dagger))
                 \nonumber \\
  &&-
  i\a_9 \,
    \str(L_{\mu\nu}D_\mu\Sigma D_\nu\Sigma^\dagger
                        +R_{\mu\nu}D_\mu\Sigma^\dagger D_\nu\Sigma)
,\end{eqnarray}
that contribute to meson form factors at tree level.
Here $L_{\mu\nu}$, $R_{\mu\nu}$ are the field-strength tensors of 
the external sources, which for an
electromagnetic source are given by
\begin{eqnarray} \label{eqn:LR}
  L_{\mu\nu} = R_{\mu\nu}
  = e\cQ(\partial_\mu \cA_\nu-\partial_\nu \cA_\mu)+ie^2\cQ^2[\cA_\mu,\cA_\nu]
.\end{eqnarray}
Unlike quenched chiral perturbation theory, where the low-energy constants are distinct from those 
in chiral perturbation theory, the partially quenched  parameters in
Eq.~\eqref{eqn:L4PQQCD} are the dimensionless Gasser-Leutwyler
parameters of chiral perturbation theory~\cite{Gasser:1985gg},
which can be demonstrated by matching.

\section{Form Factors in Infinite Volume} \label{s:infinite}

The electromagnetic form factor $G_{X}$ of an octet meson 
$\phi_X$
is required by
Lorentz invariance and gauge invariance to have the form
\begin{equation}\label{eqn:mesonff}
  \langle\phi_{X}(p')|J_\mu|\phi_{X}(p)\rangle
  = 
  e \, G_{X}(q^2)(p+p')_\mu,
\end{equation}
where
$p$ ($p'$) is the momentum of the incoming (outgoing) meson, 
and $q_\mu=(p'-p)_\mu$ is the momentum transfer.
Conservation of electric charge protects it from
renormalization,
hence at
zero momentum transfer
$e \, G_X(0)=Q_X$, where $Q_X$ is the charge of $\phi_X$.
The charge radius $r_{X}$
is related to the slope of $G_{X}(q^2)$ at $q^2=0$,
namely
\begin{equation}
  <r_{X}^2>
  =
  -6\frac{d}{dq^2}G_{X}(q^2)\Big|_{q^2=0}
.\end{equation}
Charge conjugation implies the form factor relations: $G_{\pi^+}(q^2) = - G_{\pi^-}(q^2)$, 
$G_{K^+}(q^2) = - G_{K^-}(q^2)$, and $G_{K^0}(q^2) = - G_{\overline{K}^0}(q^2)$,
as well as $G_{\pi^0}(q^2) = G_{\eta}(q^2) = 0$.

\begin{figure}[tb]
  \includegraphics[width=0.75\textwidth]{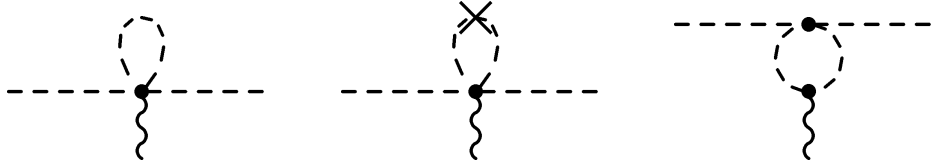}%
  \caption{
    Loop diagrams contributing to the octet meson charge radii
    in partially quenched chiral perturbation theory.
    Octet mesons are denoted by a dashed line,
    singlets (hairpins) by a crossed dashed line, 
    and the photon by a wiggly line.
  }
  \label{F:pions}
\end{figure}

\begin{figure}[tb]
  \includegraphics[width=0.50\textwidth]{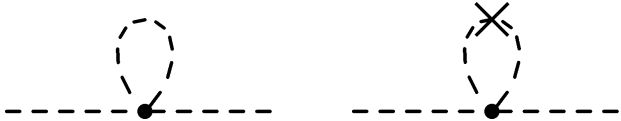}%
  \caption{
    Wavefunction renormalization diagrams  
    in partially quenched chiral perturbation theory.
  }
  \label{F:pions-wf}
\end{figure}

To calculate the charge radii to lowest order
in the chiral expansion 
one has to include operators of $\cL$ in Eq.~\eqref{eqn:Lchi}
to one-loop order 
[see Figs.~(\ref{F:pions}) and (\ref{F:pions-wf})]
and operators of Eq.~\eqref{eqn:L4PQQCD}
to tree level. 
Using dimensional regularization, where
our subtraction scheme removes 
$\frac{1}{\e}+1-\gamma_{\text{E}}+\log 4\pi$,
we find the form factors have the form
\begin{equation}
 \label{eq:answer}
G_X(q^2) 
= 
Q_X \left( 1 - \frac{4 \alpha_9}{f^2} q^2  \right)
+ 
\frac{1}{(4\pi f)^2} \sum_\phi A^X_\phi \, F(m_\phi^2, q^2)
.\end{equation}
In this general expression, $Q_X$ is the meson charge, the 
sum on $\phi$ runs over all non-degenerate loop mesons of mass $m_\phi$. 
The coefficients $A_\phi^X$ are products of charge and Clebsch-Gordan
factors for the coupling of the loop meson $\phi$ to the external state
meson $X$. These coefficients are listed in Table~\ref{t:A}. 
If a particular loop meson is not listed, then the values of $A_\phi^X$
are identically zero for all states $X$.
Notice that in the isospin limit the charges $q_j$ and $q_l$ always enter
in the combination $q_j + q_l = - q_r$ and for this reason do not explicitly 
appear in coefficients listed in the table.
Lastly the non-analytic function $F(m^2, q^2)$ above is defined to be
\begin{equation}
  F(m^2, q^2)
  = \frac{1}{6} \left[ 
  q^2 \, \log\frac{m^2}{\mu^2}
  + 4 
  m^2\, {\mathcal F}\left(\frac{-q^2}{4 m^2}\right)
  \right]
,\end{equation}
where the auxiliary function ${\mathcal F}(a)$ is given by
\begin{equation} \label{eqn:Fa}
   {\mathcal F}(a)
   = 
   \left( a- 1 \right)
   \sqrt{1-\frac{1}{a}}
   \log\frac{\sqrt{1-\frac{1}{a}+i\e}-1}{\sqrt{1-\frac{1}{a}+i\e}+1}
   +\frac{5}{3} a -2 
.\end{equation}
In the limit $m_j\to m_u$, $m_r\to m_s$ and using the 
physical quark charges, 
we recover the chiral perturbation theory result~\cite{Gasser:1985gg,Gasser:1985ux}.

\begin{table}
\caption{\label{t:A}
Coefficients $A_\phi^X$ of loop mesons contributing to electromagnetic form factors.
We list the coefficients only for $X = \pi^+$, $K^+$, and $K^0$ states because the rest follow from 
charge conjugation: $A_\phi^{\pi^+} = - A_\phi^{\pi^-}$,  $A_\phi^{K^+} = - A_\phi^{K^-}$,
and  $A_\phi^{K^0} = - A_\phi^{\overline{K}^0}$ for all $\phi$. And of course
$A_\phi^{\pi^0} = A_\phi^{\eta} = 0$. 
}
\begin{tabular}{c | c c c c }\hline\hline
  $X$           & \multicolumn{4}{c}{$\phi$} \\
  		& $\qquad ju \qquad$            & $\qquad ru \qquad$        & $\qquad js \qquad$            & $\qquad rs \qquad$  \\ \hline
  $\pi^+$ 	& $2 (q_u - q_d)$ & $q_u - q_d$ & $0$             & $0$   \\
  $K^+$         & $2 q_u + q_r$   & $q_u - q_r$ & $- 2 q_s - q_r$ & $q_r - q_s$  \\
  $K^0$         & $2 q_d + q_r$   & $q_d - q_r$ & $- 2 q_s - q_r$ & $q_r - q_s$  \\ \hline\hline
\end{tabular}
\end{table}

Using the values from the Table, the form factor of the $\pi^+$, e.g., appears as 
\begin{eqnarray}\label{eqn:donaldduck}
  G_{\pi^+}(q^2)
  &=& Q_{\pi^+} 
\left\{ 
 1 - \frac{4 \a_9}{f^2}q^2
+  \frac{1}{(4 \pi f)^2}
  \left[2 F (m_{ju}^2, q^2) + F(m_{ru}^2, q^2) \right]
\right\}
,\end{eqnarray}
where the charge of the pion is $Q_{\pi^+} = q_u - q_d$.
The pion form factor appears rather special at one loop due to its independence 
from the charges of the sea quarks.\footnote{At one loop, we can see explicitly that the 
independence from sea quark charges arises from cancellations between isospin degenerate loop mesons. 
Away from the isospin limit
the sea charges remain, e.g., in non-degenerate $SU(4|2)$ the non-vanishing loop coefficients for the $\pi^+$
are: $A^{\pi^+}_{ju} = q_u - q_j$, $A^{\pi^+}_{lu} = q_u - q_l$, $A^{\pi^+}_{jd} = -(q_d - q_j)$, 
and $A^{\pi^+}_{ld} = -( q_d - q_l)$.  
} 
In fact, this independence from sea quark charges holds non-perturbatively
and was shown using the behavior of lattice correlators under charge conjugation in~\cite{Draper:1989bp}.
We demonstrate this more simply as follows.

The electromagnetic current $J_\mu$ can be decomposed into isosinglet and isovector combinations,
$J_\mu = J^I_\mu + J_\mu^3$.
Matrix elements of the former involve the charges of the sea quarks; 
while in the isospin limit, matrix elements of
the latter are independent of these sea charges. Now consider the electromagnetic current matrix
element of the neutral pion,\footnote{%
As with the proof in~\cite{Draper:1989bp}, our argument holds for
charged $SU(2)$ mesons of arbitrary spin and parity in the isospin limit, e.g.
one can use the charge conjugation invariance of the $\rho^0$ to establish
that the form factor of the $\rho^+$ is independent of the charges of the sea.
}
\begin{equation}
\langle \pi^0(P') | \, J_\mu \, | \pi^0(P) \rangle 
=
\langle \pi^0(P') | \, J^I_\mu \, | \pi^0(P) \rangle 
+
\langle \pi^0(P') | \, J^3_\mu \, | \pi^0(P) \rangle 
.\end{equation} 
This vanishes by charge conjugation invariance. Furthermore
the isovector contribution vanishes, hence, so too must the isosinglet contribution. 
The matrix element of the isosinglet current is the same for all pions in the isospin limit. 
Thus as a consequence of isospin symmetry and charge conjugation invariance, 
we have
\begin{equation}
\langle \pi^+(P') | \, J_\mu \, | \pi^+(P) \rangle 
=
\langle \pi^+(P') | \, J^3_\mu \, | \pi^+(P) \rangle 
,\end{equation} 
and hence the charged pion form factor is independent of the sea quark charges. 

As a result of this independence from sea quark charges, 
one can efficaciously ignore operator self-contractions and still determine the 
pion form factor~\cite{Draper:1989bp}.
Note that while the pion form factor is insensitive to contributions
from closed quark loops with photon insertion, it is sensitive to sea 
quarks. As pointed out in~\cite{Arndt:2003ww},  the analogous  
calculation in quenched chiral perturbation theory shows that there is no meson mass dependence at one-loop order.
This result is also clear from the Table: only valence-sea loop mesons enter
our one-loop expressions. 
Notice that the kaon form factors are not independent from the sea quark charges. 
This dependence only disappears in the $SU(3)$ limit, which is badly violated in nature.\footnote{%
Indeed the argument presented in~\cite{Draper:1989bp} only applies for mesons consisting 
of degenerate flavors. In the non-degenerate case, the argument reduces to a demonstration that the 
sum of operator self-contractions in the meson and its charge conjugate vanish.
}

Finally we derive expressions for the charge radii at one-loop order. 
In infinite volume, we can take $q^2 \ll m^2$
to find that the meson charge radii are given by
\begin{equation} \label{eq:radius}
< r_X^2 >  
= 
Q_X  \frac{24 \alpha_9}{f^2} 
- 
\frac{1}{(4\pi f)^2} \sum_\phi A^X_\phi  
\left( \log \frac{m_\phi^2}{\mu^2} + 1
\right)
.\end{equation}

\section{Lattice Spacing Dependence} \label{s:a}

In this Section, we detail the modifications to our results at finite lattice spacing.
We consider  a mixed lattice action consisting of Ginsparg-Wilson valence
quarks and staggered sea quarks.
To address the effects of the lattice spacing, one formulates the continuum effective
theory of the lattice action and then matches this effective theory onto a chiral
perturbation theory. In this work we assume the natural hierarchy of scales
\begin{equation} \notag
m_q \ll \Lambda_{\text{QCD}} \ll \frac{1}{a},
\end{equation}
and choose the power counting
\begin{equation}
p^2 \sim m_\pi^2 \sim a^2 \Lambda_{\text{QCD}}^4 
.\end{equation}
The form factors can now be systematically calculated in the dual expansion 
in quark mass and lattice spacing.
Such modifications to the electromagnetic form factors of mesons are rather simple. 
It was demonstrated in Ref.~\cite{Arndt:2004we} that no local $a$-dependent operators
contribute to the form factors to $\mathcal{O}(p^2)$ due to charge conservation. 
In our power counting, local corrections to the current in the $a^2$ chiral Lagrangian
will contribute at $\mathcal{O}(p^4)$ and will be competitive with two-loop effects. 
These local terms can thus be neglected here. 
The lattice spacing corrections to meson form factors then enter only through
the dependence of the meson masses on the lattice spacing.

With a mixed lattice action, there is no symmetry that 
relates the valence and sea sectors of the theory. The Symanzik Lagrangian thus contains
dimension-six mixed field bilinears of the form~\cite{Bar:2003mh,Bar:2005tu}
\begin{equation} \notag
\delta \mathcal{L}^{(6)} \sim 
\left( \ol Q \, \Gamma \, P_V  Q \right)
\left( \ol Q \, \Gamma \, P_S  Q \right)
,\end{equation}
where $P_V$ is a diagonal matrix that has unit entries corresponding to the valence sector, and 
$P_S$ is a diagonal matrix that has unit entries corresponding to the sea sector (we have also implicitly 
enlarged the vector $Q$ to transform under the continuum $SU(15|3)$ that includes four tastes 
for each flavor of sea quark). 
Above $\Gamma = \gamma_\mu$, $\gamma_\mu \gamma_5$ are the only Dirac matrices allowed 
by the chiral symmetry of the valence sector and the axial symmetry of the sea sector. 
The mapping of such mixed bilinears onto operators in the chiral perturbation theory at finite lattice 
spacing produces one new operator of the form~\cite{Bar:2005tu}
\begin{equation} \label{eq:mixed}
\delta \mathcal{L} = - \frac{1}{8} a^2 C_{\text{mix}} \, \str (T_3 \, \Sigma \, T_3 \, \Sigma^\dagger ) 
,\end{equation}
where $T_3 = P_S - P_V$. 
This operator contributes to masses of mesons formed from one valence and one sea quark. 
Such valence-sea mesons are not protected from additive mass renormalization due to the mixed action's symmetry. 
Thus to $\mathcal{O}(p^2)$, the loop meson consisting of a staggered quark $Q_i$, i.e.~of flavor $Q$ 
and taste $i$, and a valence quark $Q'$ have masses given by~\cite{Bar:2005tu}
\begin{equation} \label{eq:mixedmass}
m_{Q_i Q'}^2 = \frac{\lambda}{f^2} (m_Q + m_{Q'}) + \frac{a^2}{f^2} C_{\text{mix}}
.\end{equation}
These masses are independent of the staggered quark taste and hence there is a four-fold 
taste degeneracy in the valence-sea meson loops. This four-fold degeneracy is exactly canceled
by the $\frac{1}{4}$ factor that must be inserted by hand to implement the fourth-root trick. 
Hence in a mixed action simulation, the meson form factors at $\mathcal{O}(p^2)$ are 
given by Eq.~\eqref{eq:answer} with the loop meson masses given in Eq.~\eqref{eq:mixedmass}.

\section{Volume Dependence} \label{s:finite}

On the lattice, the available momentum modes are quantized and observables calculated
thus inherit a dependence on the lattice volume. This dependence, which is inherently a 
long-distance effect, can be ascertained using chiral perturbation theory. 
We customarily choose a hypercubic box of three equal spatial dimensions $L$, 
and time dimension $T$, with $T \gg L$. With periodic boundary conditions on the quark 
fields, the available meson momenta have the form $k_\mu = (k_0, \bm{k})$, with
$\bm{k} = \frac{2\pi}{L} \bm{n}$, and $\bm{n}$ represents a triplet of integers. 
Due to the assumed length of the time direction, we treat $k_0$ as continuous.

As spontaneous symmetry breaking does not occur in finite volumes, we must be careful
also to specify that $m_\pi L > 1$ so that mesonic zero modes do not become strongly 
coupled~\cite{Gasser:1987ah}. 
This restriction ensures that the zero modes do not conspire to restore chiral symmetry. 
Provided this is the case, the standard $p$-counting of chiral perturbation theory 
remains intact and integrals over loop momenta can merely be replaced by 
corresponding sums over quantized momenta. This replacement leads to a dependence
of physical quantities on the lattice size $L$. 

Calculation of the finite volume correction to our one-loop results for meson 
form factors is straightforward. Considering pionic matrix elements of the charge density\footnote{%
We do this only for simplicity. The form factors can be extracted from any component of the current, 
and this is the one commonly chosen in lattice QCD simulations. 
The volume effects from extracting form factors from the spatial components can similarly be calculated.
} in a box, we have 
\begin{eqnarray} 
\frac{\langle\phi_{X}(p')|J_4 |\phi_{X}(p)\rangle}{p'_4 + p_4}
&=&
e \, Q_X \left( 1 - \frac{4 \alpha_9}{f^2} q^2  \right) \notag \\
&&+ 
\sum_{\phi, \, \bm{k}} \frac{e A^X_\phi}{2 f^2 L^3}  
\int_0^1 dx   
\left[\frac{1}{\sqrt{(\bm{k} + x \bm{q})^2 + m_\phi^2 + x(1-x) q^2}} 
-
\frac{1}{\sqrt{\bm{k}^2 + m_\phi^2}}
\right]. \notag
\end{eqnarray}
The finite volume shift can then be obtained using the Poisson re-summation formula
and the resulting sums can be cast into exponentially convergent forms, see, e.g.~\cite{Sachrajda:2004mi}.
We find the finite volume shift to the meson form factors
\begin{equation} \label{eq:volume}
\delta_{L}  G_X(q^2) 
= 
\frac{1}{(4 \pi f)^2}
\sum_\phi A_\phi^X
\int_0^1 dx 
\left[ \,
\mathcal{I}(x {\bm q}, m_\phi^2 + x(1-x)q^2) 
- 
\mathcal{I}(\bm{0}, m_\phi^2) \,
\right]
,\end{equation}
where we have defined
\begin{equation}
\mathcal{I}(\bm{q}, \Delta) 
= 
\int_0^\infty d\tau 
\frac{e^{-\tau \Delta}}{\tau^2} 
\left[ 
\prod_{j=1}^3 
\vartheta_3 
\left( 
\frac{q_j L}{2}, e^{-\frac{L^2}{4 \tau}}
\right)
-
1 
\right]
,\end{equation}
with $\vartheta_3(q,z)$ as the Jacobi elliptic theta function of the third kind.

\section{Extrapolations of the charge radii} \label{s:extrap}

We now consider each of the various extrapolations necessary to make physical predictions 
from lattice data for meson form factors. For simplicity we consider the pion form 
factor and this will enable us to compare with the lattice data in~\cite{Bonnet:2004fr}. 
To perform our analysis, we fix the low-energy constant 
$\alpha_9 (\mu = 1 \texttt{GeV}) = 0.0069$ by using the 
one-loop result Eq.~\eqref{eq:radius} 
along with the experimentally determined pion charge radius~\cite{Eidelman:2004wy}.

\begin{figure}[tb]
  \centering
  \includegraphics[width=0.5\textwidth]{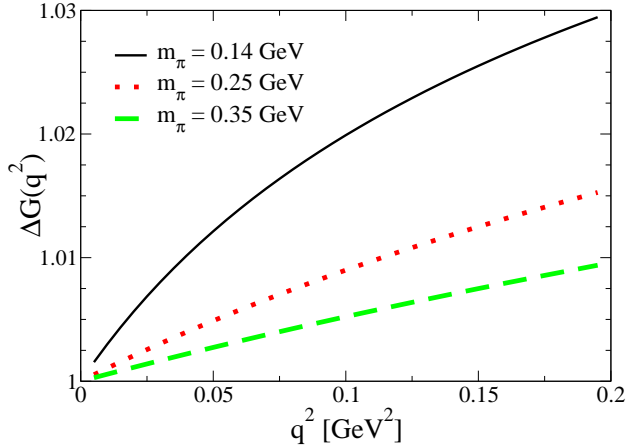}%
  \caption{Momentum transfer dependence of the pion form factor.
The ratio of the pion form factor to its slope at zero momentum transfer 
is plotted versus $q^2$ for a range of pion masses.  
  }
  \label{f:mom}
\end{figure}

\subsection{Momentum Extrapolation}

First we shall deal with the limitation of discrete lattice momenta. 
Above we have determined the effect of the finite volume on the radii 
by using the allowed lattice momenta in a periodic box. Another consequence
of periodic boundary conditions is that the momentum transfer is quantized;
hence, the limit leading to Eq.~\eqref{eq:radius} cannot be taken. Provided 
the lowest momentum transfers are in the chiral regime, the effective field 
theory can be used to perform a momentum extrapolation, \emph{cf}. Eq.~\eqref{eqn:donaldduck}. 
It is questionable whether at current lattice volumes the minimum spatial momentum  
$| \bm{q}_\text{min}| \approx 0.5 \, \texttt{GeV}$ meets this restriction. For a relativistic 
object like the pion, a slight reduction is seen because  $q^2 < \bm{q}^2$ (unlike the nucleon
where $q^2 \approx \bm{q}^2$). Furthermore, corrections 
to Eq.~\eqref{eqn:donaldduck} from higher terms in the chiral expansion 
are of order $q^2 / \Lambda_\chi^2$, for which the smallest available $q^2$ in Ref.~\cite{Bonnet:2004fr} 
yields $\sim 20 \%$ corrections. Thus we shall assume that the momentum transfer 
dependence of the lattice data at the minimal value of $q^2$ is captured by the effective 
field theory.

With this assumption, let us investigate the momentum transfer dependence of the pion 
form factor at one-loop order. If the minimal momentum transfer 
$q^2$ is indeed small compared to $m^2$, then the form factor will have linear 
behavior in $q^2$ and no momentum extrapolation is necessary
to determine the slope near $q^2 = 0$. On the other hand, imagine that $m^2$ is small compared 
to $q^2$. In this limit, the chiral logarithm dominates the form factor, which is multiplied by $q^2$,
and again the behavior is linear. In the intermediate region $q^2 \sim 4 m^2$ the behavior  
of the function $\mathcal{F}[-q^2 / (4 m^2)]$ in Eq.~\eqref{eqn:Fa} becomes important. This is the region of parameters 
relevant for current lattice simulations. In Figure~\ref{f:mom}, we plot
the function $\D G(q^2)$, defined by
\begin{equation}
\D G(q^2) = \frac{G_\pi(q^2) - Q_\pi}{q^2 G'_\pi(q^2)}
,\end{equation}
as a function of $q^2$ in  order to see the deviation from linearity. 
The largest deviation from linearity over this range of momentum transfer 
is for the physical pion mass. For the larger pion masses employed on the 
lattice, the plot shows that for all practical purposes 
we can treat the form factor as linear in $q^2$. For example, 
the data at the lowest $q^2_\text{min}$ and at lightest pion mass in Ref.~\cite{Bonnet:2004fr}, 
the difference from linearity is less than one percent. Comparatively, the neglected higher-order terms in 
the chiral expansion are more than an order of magnitude larger
(these terms, however, modify the momentum transfer dependence).

\begin{figure}[tb]
  \centering
  \includegraphics[width=0.5\textwidth]{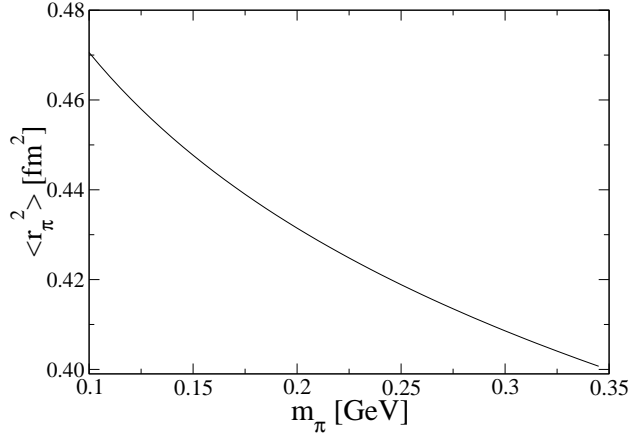}%
  \caption{Meson mass dependence of the pion charge radius. 
  }
  \label{f:chiral}
\end{figure}

\subsection{Chiral Extrapolation}

Most striking is the well-known behavior of the charge radii near the chiral limit. 
This can be seen from the chiral logarithm in Eq.~\eqref{eq:radius}.
For arbitrarily light pions the virtual cloud of pionic excitations in a meson extends arbitrarily far,
and hence there is considerable variation of the charge radius as a function of the pion mass. 
One must be careful, however, not to assume that such variation with mass will necessarily be seen in lattice data. 
Looking at Eqs.~\eqref{eq:radius}, and \eqref{eq:mixedmass}, we see that the valence-sea meson mass
is what dominates the chiral behavior at one-loop order. Thus while the valence-valence 
and sea-sea meson masses are light, the additive mass renormalization allowed by the mixed-action symmetry 
breaking may actually push one away from the chiral regime. In Figure~\ref{f:chiral}, we plot 
the meson mass dependence of the pion charge radius. Keep in mind that the meson mass 
which is relevant is the valence-sea mass. Ignoring other systematic errors, lattice calculations 
with valence-sea meson masses $\sim 350 \, \texttt{MeV}$ will undershoot the pion charge radius
by $\sim 10 \%$.

\subsection{Volume Extrapolation}

\begin{figure}[tb]
  \centering
  \includegraphics[width=0.5\textwidth]{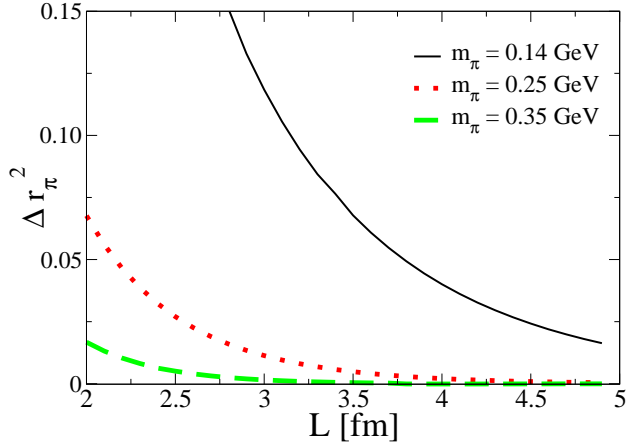}%
  \caption{Volume dependence of the pion charge radius. The relative difference $\D \, r_\pi^2$
of the charge radius in finite to infinite volume is plotted as a function of the lattice size $L$. 
The curve for $m_\pi = 0.14 \, \texttt{GeV}$ terminates for $m_\pi L = 2$ where pionic zero modes 
become important. 
  }
  \label{f:vol}
\end{figure}

The virtual pion cloud will be affected by the boundary conditions imposed
in the lattice simulation.
Due to the chiral singularity in the one-loop contribution, 
one might also expect that the volume effects are substantial for light pions
because these corrections stem from the long distance physics. 
Using Eq.~\eqref{eq:volume} we can plot the volume effect as a function of 
$L$ for various values of the (valence-sea) pion mass. In Figure~\ref{f:vol},
this is done for the quantity $\D \, r_\pi^2$ which is defined by
\begin{equation}
\D \, r_\pi^2  =  \frac{< r_\pi^2>_L}{< r_\pi^2>}
,\end{equation}
where $< r_\pi^2>$ is the infinite volume radius, which is given in Eq.~\eqref{eq:radius}, 
and $< r_\pi^2>_L = -6 \frac{d}{dq^2} \delta_L G_\pi(q^2)\big|_{q^2 = 0}$ is the finite volume
modification. We see that at the physical value of the pion mass, volume effects 
are substantial $\sim 12\%$ in a $3 \, \texttt{fm}$ box consistent with our intuition.
However, the volume effects drop considerably for larger pion masses.  For an $m = 0.35 \, \texttt{GeV}$
valence-sea pion in a $2.5 \, \texttt{fm}$ box [which roughly corresponds to the values 
used in~\cite{Bonnet:2004fr} ignoring the effects of the mixed action in Eq.~\eqref{eq:mixedmass}], 
the effects of periodic boundary conditions in a finite box lead to a 
bigger positively charged pion, but by only a negligible $0.5 \%$.

\subsection{Continuum Extrapolation}

Lastly we investigate the systematic error in the pion form factor data due to the finite 
size of the lattice spacing. 
The lattice spacing $a$ enters our expressions only through the valence-sea meson masses,
Eq.~\eqref{eq:mixedmass}. Ideally we would have data at multiple values of 
the lattice spacing and quark mass to enable a proper continuum and chiral extrapolation
of the data. In Ref.~\cite{Bonnet:2004fr}, there is only one lattice spacing employed
and only one valence-valence meson mass light enough to warrant a chiral analysis. 
As we have commented above, volume and momentum extrapolations produce negligible
corrections compared to neglected higher-order chiral contributions. Thus we need
only use the infinite volume expression in Eq.~\eqref{eqn:donaldduck} to make contact with the
data. Fixing 
$\alpha_9$ 
and 
$f$ 
to their physical values, we can estimate the 
mixed action low-energy constant 
$C_\text{mix}$.

Using the form factor data at the lowest value of 
$q_{\text{min}}^2 = 0.18 \, \texttt{GeV} \, {}^2$
with the valence pion mass 
$m_{\pi} = 0.32 \, \texttt{GeV}$ 
and the sea pion mass
$m_{jj} = 0.35 \, \texttt{GeV}$, 
we estimate
$C_\text{mix} = 0.0064 \, \texttt{GeV}\,{}^6$. 
We will not cite errors on this value
because it is a rough estimate---only one datum is used to determine 
$C_\text{mix}$. 
Moreover as it enters our expressions logarithmically, any error in the form factor exponentiates
into our estimate of 
$C_\text{mix}$. 
Using the statistical error bars from the
lattice data for a high-low estimate, we find a rather wide range statistically allowed 
for 
$C_\text{mix}$ 
from 
$-0.0040$ 
to 
$0.080$ 
in units of 
$\texttt{GeV} \, {}^6$. 
The range produced from the systematic uncertainty is comparable.

\section{Summary} \label{s:summy}

Above we have investigated various extrapolations necessary to 
connect lattice QCD data for meson form factors at small momentum 
transfer to the physical meson charge radii. 
We find that while volume effects are sizable at the physical pion mass, 
they are negligible for 
$m_\pi \sim 250$ - $350 \, \texttt{MeV}$ 
in current lattice volumes. 
Provided the minimum lattice momentum 
$q_\text{min} \lesssim 300 \, \texttt{MeV}$, 
there is no need for a momentum extrapolation: the small 
$q^2$-dependence 
predicted from chiral perturbation theory is very linear. 
The corrections to this linearity arise from next-to-next-to-leading
order contributions in the chiral expansion.
At these pion masses and momenta, we thus conclude that systematic error 
is dominated by higher-order terms in the chiral expansion: 
$\sim 10\%$ from corrections that scale generically as $m_\pi^2 / \L_\chi^2$, 
and $\sim 20\%$ from terms of order $q^2/\L_\chi^2$ with $q^2 = q^2_{\text{min}}$
of~\cite{Bonnet:2004fr}.
For the mixed action data in~\cite{Bonnet:2004fr}, there are additional 
errors from lattice spacing artifacts. Demanding that their 
data be described within the effective field theory, 
we are able to estimate the low-energy constant 
$C_\text{mix}$ 
that parametrizes the
explicit breaking of the mixed action symmetry 
[$SU(6|3) \to SU(3|3) \otimes SU(3)$] 
at 
$\cO(a^2)$. 
We obtained a rough value 
$C_\text{mix} \approx 0.006 \, \texttt{GeV}\,{}^6$, 
where there 
is sizable error in this value arising from the statistical error in the lattice data
and systematic error due to neglecting of two-loop results (for both the quark mass
and momentum dependence of the pion form factor). Because 
$C_\text{mix}$ 
enters the expression for the form factor through a logarithm, a small 
$\sim 10 \%$
statistical or systematic error in the lattice data results in a very large 
uncertainty for 
$C_\text{mix}$. 
Data at multiple lattice spacings or with varying quark mass 
would be ideal for a better determination of this low-energy constant.
Nonetheless, the pion charge radius is a quantity that is sensitive
to the fermion discretization, and can be used to understand the errors 
associated with the continuum extrapolation of mixed-action lattice data.

The computation of meson form factors on the lattice is challenging. 
For the pion, however, the absence of operator self-contractions
in the isospin limit~\cite{Draper:1989bp}, puts the charge radius in reach of 
current lattice technology.
Further data at additional values of the valence quark mass and lattice spacing
will thus allow one to predict the pion charge radius at about the 
$\sim 10$--$20\%$ 
level or better, without any model assumptions.

\begin{acknowledgments}
We would like to thank A.~Walker-Loud for initial involvement, and 
R.~Lewis for providing the lattice data from~\cite{Bonnet:2004fr}. 
B.C.T.~acknowledges the Institute for Nuclear Theory at the University of 
Washington for hospitality and partial support during the completion of this work.
This work is supported in part by the U.S.\ Department of Energy under 
Grant No.\ DE-FG02-05ER41368-0.
\end{acknowledgments}

\appendix

\bibliography{hb}

\end{document}